\documentclass[11pt]{article}

\usepackage[]{graphics,graphicx}

\usepackage{amsmath}
\usepackage{pstricks}

\usepackage{blindtext}

\usepackage{setspace} 

\RequirePackage{lineno}

\begin{document} 
\modulolinenumbers[1]


\begin{center}
{\LARGE\bf Kinematics of erosion-induced landslide mobility}
\end{center}
\vspace{7mm}
{Shiva P. Pudasaini
\\[3mm]
Technical University of Munich, 
School of Engineering and Design\\
Civil and Environmental Engineering}\\
{Arcisstrasse 21, D-80333, Munich, Germany}\\[3mm]
Kathmandu Institute of Complex Flows\\ Kageshwori Manohara - 3,
Bhadrabas, Kathmandu, Nepal\\[2mm]
{E-mail: shiva.pudasaini@tum.de}\\[7mm]
\noindent
\noindent
{\bf Abstract:}
Erosion-induced landslide mobility is a prominent phenomenon in nature. However, this topic remains poorly understood, even though erosion greatly alters nearly every aspect of landslide mechanics, dynamics, run-out, mobility, deposition structure and the enormous destructive power carried by such catastrophic events. Here, I present a simple, elegant, and exact analytical model describing the kinematics of erosion-induced landslide mobility in terms of a newly constructed mobility controller containing essential forcing mechanisms in erosion. In contrast to the classical perspective, I formally prove that, irrespective of how large the erosion rate is, the erosion rate alone is not sufficient to determine enhanced or reduced mobility; mobility is also greatly regulated by the volume bulking rate, erosion drag or thrust, and flow depth. With this mechanical mobility controller, it is now possible to explain precisely when the mobility of an erodible landslide is enhanced or reduced. I demonstrate that, a mechanically valid, physically founded erosion model must be able to explain both the enhanced and reduced mobility of erosive landslides.

\section{Introduction}

As a complex mechanical process, erosion  dominantly controls the dynamics of geophysical mass flows including landslides, avalanches and debris flows and plays an important role in mass transport and evolution of the mountain landscape (Huggel et al., 2005; Hungr et al., 2005; Santi et al., 2008; Evans et al., 2009; de Haas et al., 2020; Mergili et al., 2020). As these flows cascade down mountain slopes the sediment is entrained from the bed resulting in disproportionately increased volumes and destructive powers by several orders of magnitude and become exceptionally mobile (Theule et al., 2015; Cuomo et al., 2016; Somos-Valenzuela et al., 2016, Mergili et al., 2018; Liu et al., 2019; de Haas et al., 2020). Erosion-induced excessive volume is a key control on the flow dynamics including the flow velocity, depth, travel distance and impact area (Le and Pitman, 2009; Dowling and Santi, 2014; Pudasaini, 2025). So, erosion, entrainment and associated flow bulking in landslide prone areas and debris-flow torrents are a major concern for earth scientists, civil and environmental engineers and landuse planners (Huggel et al., 2005; Evans et al., 2009; Mergili et al., 2020). Mobility is among the most important features of the erosive landslide as it directly measures the threat posed by the landslide. Landslide mobility is characterized by an enormous impact force, exceptional travel distance and inundation area.
\\[3mm]
Recently, there has been a rapid increase in the studies of erosion and entrainment associated with devastating mountain mass transports (Egashira et al., 2001; Fraccarollo and Capart, 2002; McDougall and Hungr, 2005; Le and Pitman, 2009; Christen et al. 2010; Iverson and Ouyang, 2015; Frank et al., 2016). However, these models are mechanically inconsistent, because they do not include the erosion-induced net momentum production. Pudasaini and Fischer (2020) proposed a process-based erosion model. Their mechanical erosion model proved that the effectively reduced friction (force) in erosion is equivalent to the momentum production. This shows that erosion can enhance the mass flow mobility. The Pudasaini and Fischer (2020) model built a foundation for erosive mass flows by mechanically including the momentum production into the momentum balance equation.
\\[3mm]
Extending the Pudasaini and Fischer (2020) model, Pudasaini and Krautblatter (2021) addressed the important issue of erosion-induced landslide mobility by explicitly deriving mechanical conditions for the mobility of erosive landslides in terms of the erosion velocity of the mobilized bed material. They mechanically explained how and when erosive landslides enhance or reduce their mobility. This was made possible by physically correctly considering the reduced inertia and the induced momentum production of the erosive landslide. This model distinctly quantifies the mobility of an erosive landslide. Based on the Pudasaini and Fischer (2020) and Pudasaini and Krautblatter (2021) models, Pudasaini (2025) further developed the first-ever unified, consistent and complete mechanical erosion model for multi-phase mass flows. These physics-based advanced models (Pudasaini and Fischer, 2020; Pudasaini and Krautblatter, 2021; Pudasaini, 2025) constitute a foundation for physically meaningful simulation of landslide motion with erosion and have clearly indicated the need of the mechanical erosion rates and the erosion-induced net momentum productions. The importance of the  mechanical erosion model for multi-phase mass flows is widely realized, and it is applied in simulations of real catastrophic multi-phase events (Li et al., 2019; Pudasaini and Mergili, 2019; Qiao et al., 2019; Shen et al., 2019; Liu and He, 2020; Mergili et al., 2020; Liu et al., 2021; Zhu et al., 2026).
\\[3mm]
As an independent, alternative proof that erosion can result in enhanced or reduced mobility of erosive landslides proposed by Pudasaini and Krautblatter (2021), here, I present a novel, exact analytical model describing the kinematics of erosion-induced landslide mobility. This is made possible by a newly constructed mobility controller containing essential forcing mechanisms in erosion. In contrast to the prevailing perception, it clearly appears that, the erosion rate alone is not sufficient to determine the enhanced or reduced mobility. The mobility is also vigorously ruled by other intrinsic mechanisms and flow dynamics - the volume bulking rate, the drag or thrust, and the flow depth. The mobility controller explains exactly when and how the mobility of an erodible landslide is enhanced or reduced. With this, I formally prove that, if the erosion model is physically based, it must explain both the enhanced and reduced mobility of erosive landslides.

\section{The model construction}

Consider mass balance equations in the reference non-erosive, and erosive landslides down a channel (Pudasaini, 2025):
\begin{equation}
\displaystyle{\frac {\partial h}{\partial t} + \frac {\partial (hu)}{\partial x} = 0},
\\[1mm]
\label{Eqn_01}
\end{equation}
\begin{equation}
\displaystyle{\frac {\partial h_e}{\partial t} + \frac {\partial (h_eu_e)}{\partial x} = E},
\label{Eqn_02}
\end{equation}
where $t$ [s] is time, $x$ [m] the distance along the slope, $h$ [m] is the flow depth, $u$ [ms$^{-1}$] the mean flow velocity, and $E$ [ms$^{-1}$] is the erosion rate. In (\ref{Eqn_02}), the subscript $_e$ indicates the quantities associated with erosion, namely, the flow depth $h_e$ and the velocity $u_e$. Assume that there exists a functional relation between $h_e$ and $h$: $h_e = \mathcal H(h)$.
Then, from (\ref{Eqn_01}) and (\ref{Eqn_02}):
\begin{equation}
\displaystyle{\frac {\partial h}{\partial t} = -\frac {\partial (hu)}{\partial x}},
\label{Eqn_01aa}
\end{equation}
and,
\begin{equation}
\displaystyle{\frac {\partial (h_eu_e)}{\partial x}
= E - \frac {\partial h_e}{\partial t}
= E - \frac {\partial \left(\mathcal H(h)\right)}{\partial t}
= E - \frac{\partial\left( \mathcal H(h)\right)}{\partial h}\frac{\partial h}{\partial t}
= E + \mathcal H'(h)\frac{\partial (hu)}{\partial x}
}.
\label{Eqn_02aa}
\end{equation}
So, (\ref{Eqn_01}) and (\ref{Eqn_02}) can be combined into a single equation, yielding:
\begin{equation}
\displaystyle{h_e\frac {\partial u_e}{\partial x} + \left( \frac {\partial h_e}{\partial x}\right)u_e = E + {\mathcal H'(h)} \frac {\partial (hu)}{\partial x}},
\label{Eqn_03}
\end{equation}
where $\mathcal H' = \partial {\mathcal H}/\partial h = (\partial h_e/\partial x)/(\partial h/\partial x)$ is the volume bulking rate, a unique internal variable for erosive mass flow arching the erosive flow depth $h_e$ with the non-erosive flow depth $h$. As it is the measure of how the depth (or, gradient) of the erosive mass flow changes with the depth of the non erosive mass flow, I call $\mathcal H'$ the mobility booster. This will be clearer below. More on this is elaborated at Section 6 while discussing the results.
 With the definitions:
\begin{equation}
\displaystyle{p(x) = h_e(x),\,\, q(x) = \frac{\partial h_e}{\partial x},\,\, r(x) = E +  {\mathcal H'(h)} \frac {\partial (hu)}{\partial x}},
\label{Eqn_03a}
\end{equation}
and the associated dimensions [m], [-], [ms$^{-1}$], (\ref{Eqn_03}) can be cast in a simple, elegant form as:
\begin{equation}
\displaystyle{p(x)\frac {\partial u_e}{\partial x} + q(x)u_e = r(x)},
\label{Eqn_04}
\end{equation}
which is a first order linear ordinary differential equation of fundamental importance, see Section 3.
\\[3mm]
The presence of the term associated with $\mathcal H'$ in (\ref{Eqn_03}), or (\ref{Eqn_04}) is of great importance. In (\ref{Eqn_03}), $\displaystyle{{\mathcal H'(h)} \frac {\partial (hu)}{\partial x}}$ implicitly inherits the time-dependent flux gradient with erosion, that would not exist for steady-state. In steady-state, the flux gradient would merely be $E$, which would not tell us if the flux would be increasing or decreasing, equivalently, that would not tell us if the mobility would be increased or decreased. From the structure of (\ref{Eqn_03}) and (\ref{Eqn_04}), the role of the booster $\mathcal H'$ is clear as it amplifies the flux gradient $\frac{\partial (hu)}{\partial x}$ of the non-erosive flow to enhance the mobility with erosion.
\\[3mm]
{\bf The kinematics of erosion-induced landslide mobility:}
I call (\ref{Eqn_04}) the kinematics of erosion-induced landslide mobility.
As seen in (\ref{Eqn_04}), the mobility dynamics is explained in terms of the erosion-rate, flow depth in erosion, mass flow rate and flow depth in non-erosive flow, and how the flow depth in erosion is related to the flow depth without erosion. Equivalently, the mobility is explained with $(h, u)$ and $(E, h_e)$. The enhanced or reduced mobility is made clearer below.
\\[3mm]
The model (\ref{Eqn_04}) is generically updated at real time. This is so, because, for a given time $t$, the quantities $(h, u)$ and $(E, h_e)$; equivalently $\{p, q, r\}$; are assumed to be known or can be obtained. At each time $t$, as $\{p, q, r\}$ take unique values, (\ref{Eqn_04}) provides the velocity of erosive mass flow along the slope for that time instant. And, this process continues for all the time slices as far as erosion is active. Hence, (\ref{Eqn_04}) is an evolution equation for velocity of erosive mass flow embedding the temporal vitality.
\\[3mm]
{\bf Erosion drag-thrust dynamics:}
 The dynamics of $u_e$ in (\ref{Eqn_04}) is primarily driven by $r$, which constitutes the source, or the system forcing. In other words, if the flux gradient in the non-erosion is negligible, then, mainly the erosion rate $E$ drives the dynamics. For sufficiently large value of $E$ (and $\mathcal H'$), $u_e$ increases, otherwise $u_e$ decreases. The term associated with $q$ plays two different roles. As $q$ is the gradient of the flow depth in erosion, locally, it can take positive or negative values. If $q$ is positive, then, the term $q u_e$ acts as the drag; the velocity dependent dissipative force that becomes clearer when this term is put on the right hand side of (\ref{Eqn_04}); and the velocity $u_e$ decreases as compared to that without this term. However, if $q$ is negative, then, this leads to an erosion-thrust as it propels the flow, resulting in the enhanced velocity as compared to that without this term. So, formally, $q$ is the erosion-drag coefficient or the erosion-thrust coefficient.
This means, depending on the local gradient (the sign and magnitude) of the erosive flow depth, the dynamics may change from enhanced mobility to reduced mobility, or vice versa. This also suggests the positive contribution to mobility enhancement in the downstream from the apex of the frontal surge head of the landslide, similarly, the negative contribution to mobility enhancement in the upstream from that apex. Such a unique drag-thrust dynamics is observed here for the first time explaining the mobility of erosive-mass transport. Often in experiments, concave head structures are developed tending to reduce the mobility behind the surge apex if such surge develops.
\\[3mm]
{\bf The mobility criterion:}
 To better visualize the complex dynamics, I re-write (\ref{Eqn_04}) in an alternative form, explicitly in terms of the velocity gradient, as:
\begin{equation}
\displaystyle{\frac {\partial u_e}{\partial x} = \frac{1}{p(x)}\left[r(x) - q(x)u_e\right]}.
\label{Eqn_04a}
\end{equation}
 Whether the flow velocity in erosion will be enhanced or reduced (compared with that without erosion) depends on how $\left[r(x) - q(x)u_e\right]/p(x)$ behaves, where, as a scaling factor, $p$ amplifies enhanced or reduced flow mobility. This may also depend on whether the initial velocity is below, or above the asymptotic limit the system can approach far downstream. In realistic flow situation, the initial velocity is below the system can reach that limit. If so, the fundamental dynamics is determined by the competition between $r$ (primarily the erosion rate $E$) and $q$, which is the gradient of the flow depth of the erosive landslide, but also on $\mathcal H'$, the flux gradient in none-erosion (generally a positive quantity for rapid flows on slopes) and the flow depth $p$.
In usual situation, as the erosion rate $E$ amplifies, the flow depth increases, so is $\mathcal H'$. This means, already, there are complex relationships between the erosion rate, the flow depth, its gradient, the volume bulking rate $\mathcal H'$ (with erosion) with respect to the non-erodible flow depth and the flux gradient determining the strengths of $p, q$ and $r$ in (\ref{Eqn_04a}) as they are interrelated.
\\[3mm]
It is important to perceive that whether erosion enhances or reduces the landslide mobility must be determined by simultaneously analysing its dynamical behaviour with reference to the dynamics without erosion.
Then comes the major role played by $\left[r(x) - q(x)u_e\right]/p(x)$ and how the dynamics is amplified. The main criterion should be able to explain whether or not the velocity with erosion surpasses the velocity without erosion.
Depending on the state of $\{p, q, r\}$, the erosion-induced mobility is enhanced if $u_e > u$, reduced if $u_e < u$, and may remain neutral if $u_e \approx u$. This means, the sigh of $\left[r(x) - q(x)u_e\right]$ is not a measure of enhanced or reduced mobility. This becomes clearer when presenting the results at Section 5.

\section{The universal physics and engineering carried by (\ref{Eqn_04})}

The model equation of type (\ref{Eqn_04}) is rich as it describes wide variety of physical and engineering problems in nature. It represents some universal, fundamental processes found across many branches of science and engineering. It operates as a system responding to an external force ($r$) while being constrained by its own internal dynamics ($q$) scaled by inertia ($p$). In this system, there are four important aspects: First, $u_e$ is the state variable. In general, it can be the velocity of a moving substance, temperature along a heated rod, chemical concentration, pollutant transport in a river, or the voltage in a circuit. Second, $\frac{\partial u_e}{\partial x}$ is the rate of change of $u_e$. Third, $r$ is the driving force (or, source) of the system as an external input supplying energy (could also be mass, or force). Fourth, $p$ and $q$ are physical properties of the system. They define the scaling and resistance (or, thrust) controlling  the system reaction.
\\[3mm]
The first, second and the third terms in (\ref{Eqn_04}) are the resistance to change (inertia), decay (or, self-regulation), and the external energy (or, the net momentum production). The physical mechanism is as follows: the external push $r$ applied to a system is balanced by two internal processes (responses): inertia of the current state ($p\frac{\partial u_e}{\partial x}$, represents convection, advection, or driven transport), and the way the system dissipates (or, restores) energy based on its current capacity ($q u_e$, where $q$ is reaction rate, absorption coefficient, or friction factor: this is the local interaction, and describes how the state (substance) interacts with its environment in proportion to the current state). Positive $q$ implies friction, decay or absorption; negative $q$ implies self-generation (thrust).
\\[3mm]
The appearance of $q u_e$ is of particular importance as rates in many physical processes are proportional to the amount present, e.g., heat loss is proportional to temperature difference, light absorption is proportional to current intensity, radioactive material decays proportional to the remaining amount, chemical reactions consume reactant proportional to concentration.
\\[3mm]
Moreover, $r$ is independent external source or sink to the system pumping the relevant quantity (substance) in or sucking it out at position $x$, entirely regardless of how much $u_e$ is already there. Chemicals drips being injected at a specific location in a fluid flow is an example.
\\[3mm]
So, the rate at which $u_e$ changes with distance equals driving force minus resistance, scaled by $p$, in which $r$ tends to push $u_e$ to its higher level, $q u_e$ tries to pull it down (or, push it up), and $p$ determines how quickly the profile can be amplified.
Thus, in physics and engineering, equations of the form (\ref{Eqn_04}) are wide spread as they describe how a system responses to its environment. Physically, this type of equation often describes a quantity that is transported through space while simultaneously being driven toward a local equilibrium. Examples include: room heating, concentration of a substance, leaking pipe, light passing through fog, heat carried by flowing fluid, chemical concentration in blood vessels.

\section{Exact analytical solution for enhanced mobility}

Assuming that $\{p, q, r\}$ can be measured or are known, by considering the integrating factor $I(x) = e^{\int{\frac{q}{p}\,dx}}$, following the standard mathematical procedure of solving ordinary differential equations, the model (\ref{Eqn_04}) can be solved (integrated) exactly for the velocity of the landslide with erosion, yielding:
\begin{equation}
\displaystyle{u_e = {e^{-\int{\frac{q}{p}\,dx}}}\left[\int{\frac{r}{p}e^{\int{\frac{q}{p}\,dx}}dx} + c\right]},
\label{Eqn_05}
\end{equation}
where, $c$ is a constant of integration (with the dimension of [ms$^{-1}$]) which can be fixed with the physical boundary (initial) condition. As discussed above, the solution behaviour depends on $p, q$ and $r$, but also on the initial (or, boundary) velocity (at the erosion none-erosion interface). Mathematical structure of (\ref{Eqn_05}) tells that, with erosion, if the combined behaviour of the dynamic control (via $\{p, q, r\}$) remains unchanged, the velocity increases (or, decreases) monotonically as characterized by the exponential functions involved in (\ref{Eqn_05}). This will be clearer below while discussing the solution results.

\section{Explaining the erosion-induced enhanced mobility}

Without loss of generality, for simplicity, consider $\{p, q, r\}$ as constant. Mobility enhancement demands that with increasing $r$ the numerator of (\ref{Eqn_04a}) dominates the denominator in some way. For ease of explaining the main aspects of the kinematics of erosion (\ref{Eqn_04}), keeping the values of $p$ and $q$ fixed as 0.1 and 0.07, respectively, results are analyzed by varying the values of $r = [0.02, 0.0175, 0.016, 0.012]$, say, mainly the erosion rate. These are probable erosion rates as obtained from the laboratory experiments or the field data, also adopted here for better and wide distinction between different mobility  scenarios. It is important to note that whether the mobility is enhanced or reduced must always be measured with respect to the corresponding dynamical quantities, e.g., the flow velocity, with and without erosion.
\begin{figure}[t!]
\begin{center}
\includegraphics[width=13.5cm]{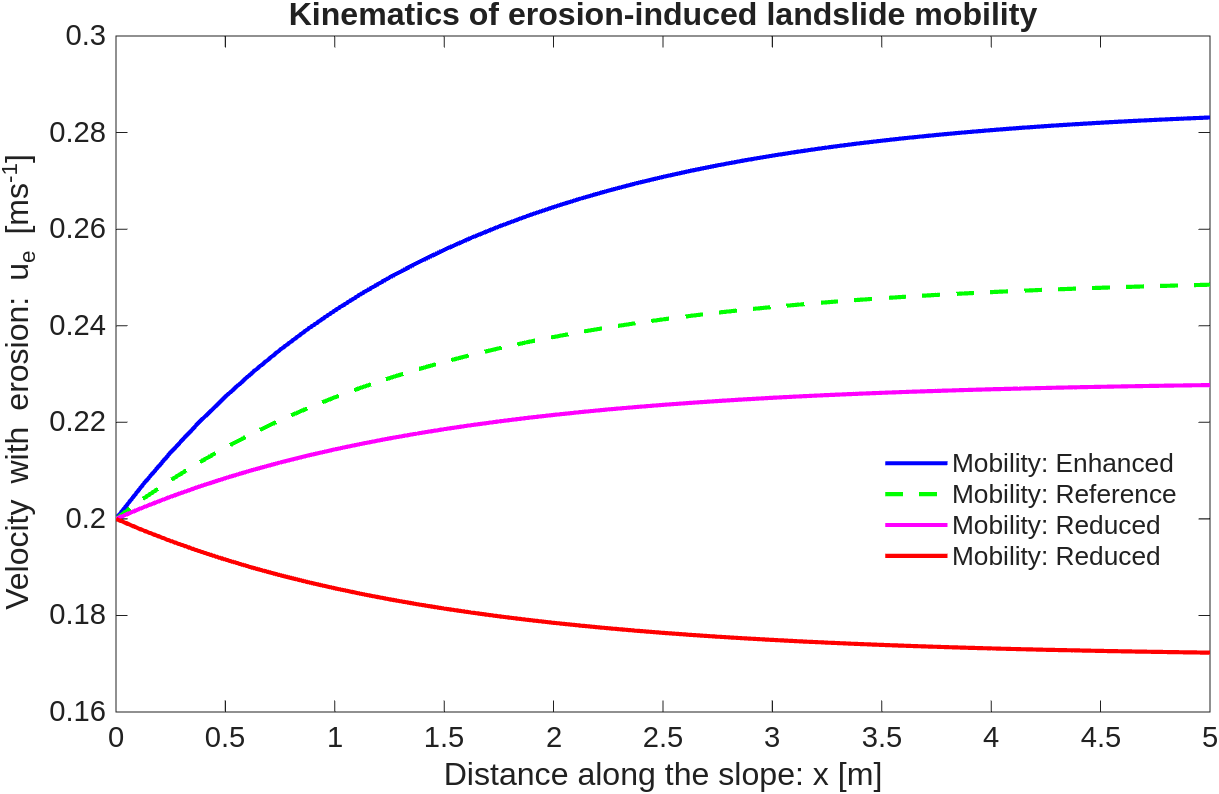}
  \end{center}
  \vspace{-5mm}
  \caption[]{Kinematics of erosion-induced enhanced ($r = 0.02$) or reduced ($r = 0.016, 0.012$) landslide mobility. The mobility is measured with the reference velocity without erosion (equivalently $r = 0.0175$), the curve in dashed green, which can also be viewed as the neutral mobility. The erosion region begins at $x = 0$.}
  \label{Fig_1}
\end{figure}
\\[3mm]
Results are presented in Fig. \ref{Fig_1}, displaying plausibly different dynamical scenarios. The reference curve in dashed green is the neutral mobility $(u_e \approx u)$ (reference) associated with $r = 0.0175$. This effectively means that there may exist a set of values of $\{p, q, r\}$ conceivably yielding a neutral state. Any solution above this describes the enhanced mobility, here, the curve in blue with $r = 0.02$, displaying the largely enhanced mobility as measured with the significantly increased velocity compared to the reference velocity without erosion. In other spectrum, any solution (increasing or decreasing) below the reference velocity (dashed green line), represents the reduced mobility, here, the lines in magenta and red with substantially smaller values of $r = [0.016, 0.012]$. Values of $r$ in between 0.02 and 0.012 (or, even exterior) provide a wide spectrum of mass flow mobility.
\\[3mm]
However, if locally the dynamic control among $p, q, r$ fluctuate and these quantities vary with their greater definitions at (\ref{Eqn_03a}), then, the velocity curves may produce locally different behaviours, even leading to complex wavy structures providing more general picture of enhanced or reduced mobility with erosion. This needs to be investigated with laboratory experiments and/or field data.

\section{Discussion}

Now, some crucial aspects of the proposed kinematics of erosion-induced landslide mobility are discussed.

\subsection{The relationship between mobility enhancement and erosion velocity}

It is important to note that the higher erosion rate is associated with the mechanically (relatively) weaker erodible bed (as defined by the inertial mass) whereas the lower erosion rate results from the mechanically (relatively) stronger erodible bed. These are reflected in the erosion velocity, the velocity of the eroded material from the bed (Pudasaini, 2025), resulting in the enhanced mobility with erosion. Hence, in principle, whether the mobility of erosive mass flows would be enhanced or reduced depends on the competition between the strength of inertial mass in the flow and the inertial mass in the erodible bed that governs the state of $\{p, q, r\}$.

\subsection{The mobility controller}

Following the results presented at Section 5, I introduce a novel concept of mobility controller in erosion. For this, re-write (\ref{Eqn_04a}) as
\begin{equation}
\displaystyle{\frac {\partial u_e}{\partial x} = \mathcal M},
\label{Eqn_04b}
\end{equation}
where, $\mathcal M = \left[r - qu_e\right]/p$. I call $\mathcal M$ the mobility controller.
As the erosion rate $E$ increases, the flow depth $p$ increases. Then, if $\mathcal M$ is increasing with $E$, the mobility is enhanced. This is the scenario when $\mathcal M$ increases with $E$ as $\mathcal H'$, the mobility booster (the volume bulking rate) increases rapidly. Following the definitions from (\ref{Eqn_03a}), a detailed inspection of $\mathcal M$ reveals that the dynamical forces are involved implicitly in $E$, $\mathcal H', h_e$ and $\partial h_e/\partial x$; of which $E$ and $\mathcal H'$ positively, $(\partial h_e/\partial x) u_e$ negatively (could also be positively) as $h_e$ normalizes the mobility. This means, there exists a set of the basic forcings and controls $\{p, q, r\}$ such that the state of $\mathcal M$ results in $\frac{\partial u_e}{\partial x} > \frac{\partial u}{\partial x}$, which essentially implies the mobility enhancement, i.e., $u_e > u$.
In other spectrum, there exists a set of $\{p, q, r\}$ such that the state of $\mathcal M$ results in $\frac{\partial u_e}{\partial x} < \frac{\partial u}{\partial x}$ implying the mobility reduction, i.e., $u_e < u$.
A closer look at Fig. \ref{Fig_1} manifests such fascinating mechanism of mobility: (I) $\frac{\partial u_e}{\partial x}$ for the curve in blue $> \frac{\partial u}{\partial x}$ for the curve in green (mobility enhanced), (II) $\frac{\partial u_e}{\partial x}$ for the curves in magenta and red $< \frac{\partial u}{\partial x}$ for the curve in green (mobility reduced). Thus, mechanically, the mobility controller $\mathcal M$ regulates the enhanced or reduced mobility of landslide with erosion.

\subsection{Mechanical-dynamical implications}

There are several important mechanical, dynamical aspects associated with the mobility controller $\mathcal M$ in (\ref{Eqn_04b}). In contrast to the classical perspective, irrespective of how large the erosion rate is, the erosion rate alone is not enough to determine the enhanced or reduced mobility, the mobility is also heavily modulated by the volume bulking rate (or the mobility booster) $\mathcal H'$, the sign and magnitude of drag (or, thrust) coefficient $q$, and the flow depth $p$. The erosion rate can be high, so is the flow depth. Yet, if the situation is such that $q > 0$, but still the volume bulking rate $\mathcal H'$ is small, then, the mobility is reduced, even in erosion. However, if $\mathcal H'$ is large enough to surpass the dissipations produced by $p$ and $q$, then, the mobility of the erosive landslide is enhanced. This means, the rate of volume bulking, $\mathcal H'$, is an immensely powerful mobility booster.
Such a clear and natural mobility enhancement mechanism is presented here for the first time with great mechanical-dynamical implications in properly understanding erosive mass transport. Yet, this needs to be extensively explored analytically and experimentally.

\subsection{Physically founded description of erosion-induced landslide mobility}

Figure \ref{Fig_1} demonstrates that depending on the mechanical state of the mobility controller $\mathcal M$, the mobility of  erosive mass flow can be enhanced or reduced (may even remain virtually unchanged).
The results presented above independently support the mechanical principle proposed by Pudasaini and Krautblatter (2021) on the erosion enhanced landslide mobility.
Consequently, I prove that models predicting only reduced mobility are mechanically invalid and physically unsound, as they fail to capture the full spectrum of erosion dynamics in natural landslides.

\section{Summary}

Erosion-induced landslide mobility is one of the most important phenomena that fundamentally alter almost every aspect of the landslide mechanics, its dynamics, run-out and the enormous destructive power carried by such disastrous natural events. Here, I presented a simple analytical model explaining the kinematics of erosion-induced landslide mobility. This was achieved by constructing a novel mobility controller in terms of essential forcing mechanisms. Then, I formally proved that the erosion rate alone is not sufficient in determining the enhanced or reduced mobility. Yet, the mobility is strongly regulated by other aspects of the flow, namely, the volume bulking rate, the drag or thrust, and the flow depth. The new model explains precisely when the mobility of an erodible landslide is enhanced or reduced. Based on the mobility controller, this provides independent support for how erosion can result in enhanced or reduced landslide mobility. This essentially implies that a physically founded erosion model must be able to explain both the enhanced and reduced mobility of erosive landslides.
\\[3mm]
{\bf Acknowledgements:} The author acknowledges the financial support from the German Research Foundation (DFG) through the research project: Landslide mobility with erosion: Proof-of-concept and application - Part I: Modeling, Simulation \& Validation; Project number 522097187.


\begin{thebibliography}{99}

{\small

\bibitem{}
 Christen, M., Kowalski, J., Bartelt, P. (2010): RAMMS: Numerical simulation of dense snow avalanches in three-dimensional terrain. Cold Reg. Sci. Technol. 63, 1-14.
\\[-5mm]
\bibitem{}  Cuomo, S., Pastor, M., Capobianco, V., Cascini, L. (2016): Modelling the space time evolution of bed entrainment for flow-like landslides. Eng. Geol. 212, 10-20.
\\[-5mm]
\bibitem{} de Haas, T., Nijland, W., de Jong, S.M., McArdell, B.W. (2020): How memory effects, check dams, and channel geometry control erosion and deposition by debris flows. Sci. Rep.10, 14024.
\\[-5mm]
\bibitem{}  Dowling, C.A., Santi, P.M. (2014): Debris flows and their toll on human life: a global analysis of debris-flow fatalities from 1950 to 2011. Nat. Hazards 71, 203-227.
\\[-5mm]
\bibitem{}   Egashira, S., Honda, N., Itoh, T. (2001): Experimental study on the entrainment of bed material into debris flow. Phys. Chem. Earth (C.) 26, 645-650.
\\[-5mm]
\bibitem{} Evans, S.G., Bishop, N.F., Smoll, L.F., Murillo, P.V.,
Delaney, K.B., Oliver-Smith, A. (2009): A re-examination of the mechanism and human impact of catastrophic mass flows originating on Nevado Huascaran, Cordillera Blanca, Peru in 1962 and 1970. Eng. Geol. 108,96-118.
\\[-5mm]
 \bibitem{}  Fraccarollo, L., Capart, H. (2002): Riemann wave description of erosional dam-break flows. J. Fluid Mech. 461, 183-228.
\\[-5mm]
\bibitem{}  Frank, F., McArdell, B. W., Huggel, C.  Vieli, A. (2015): The importance of entrainment and bulking on debris flow runout modeling: examples from the Swiss Alps. Nat. Hazards Earth Syst. Sci.15, 2569-2583.
\\[-5mm]
\bibitem{} Huggel, C., Zgraggen-Oswald, S., Haeberli, W., K\"a\"ab, A., Polkvoj, A.,
Galushkin, I., Evans, S.G. (2005): The 2002 rock/ice avalanche at Kolka/Karmadon, Russian Caucasus: assessment of extraordinary avalanche formation and mobility, and application of QuickBird satellite imagery. Nat. Hazards Earth Syst. Sci. 5, 173-187.
\\[-5mm]
\bibitem{} Hungr, O., McDougall, S., Bovis, M. (2005): In Debris-Flow Hazards and Related Phenomena(eds. Jakob, M. \& Hungr, O.) (Springer, Berlin).
\\[-5mm]
\bibitem{} Iverson, R.M., Ouyang, C. (2015): Entrainment of bed material by earth-surface mass flows: review and reformulation of depth-integrated theory. Rev. Geophys. 53, 27-58.
\\[-5mm]
\bibitem{}   Le, L., Pitman, E.B. (2009): A model for granular flows over an erodible surface. SIAM J. Appl. Math.70, 1407-1427.
\\[-5mm]
\bibitem{} Li, P., Shen, W., Hou, X., Li, T. (2019): Numerical simulation of the propagation process of a rapid flow-like landslide considering bed entrainment: a case study. Eng. Geol.263, 105287.
\\[-5mm]
\bibitem{}  Liu, W., He, S. (2020): Comprehensive modelling of runoff-generated debris flow from formation to propagation in a catchment. Landslides. https://doi.org/10.1007/s10346-020-01383-w.
\\[-5mm]
\bibitem{} Liu, W., Yang, Z., He, S. (2021): Modeling the landslide-generated debris ﬂow from formation to propagation and
run-out by considering the eﬀect of vegetation. Landslides 18,43-58.
\\[-5mm]
\bibitem{}  Liu, W., Wang, D., Zhou, J. He, S. (2019): Simulating the Xinmo landslide runout considering entrainment effect. Environ. Earth Sci.78, 585.
\\[-5mm]
\bibitem{} McDougall, S., Hungr, O. (2005): Dynamic modelling of entrainment in rapid landslides. Can. Geotech. J. 42,
1437-1448.
\\[-5mm]
\bibitem{} Mergili, M., Jaboyedoff, M., Pullarello, J., Pudasaini, S.P. (2020): Back calculation of the 2017 Piz Cengalo - Bondo landslide cascade with r.avaflow: what we can do and what we can learn. Nat. Hazards Earth Syst. Sci. 20, 505-520.
\\[-5mm]
\bibitem{} Mergili, M., Emmer, A., Juricova, A., Cochachin, A., Fischer, J.-T.,
Huggel, C., Pudasaini, S.P. (2018): How well can we simulate complex hydro-geomorphic process chains? The 2012 multi-lake outburst flood in the Santa Cruz Valley (Cordillera Blanca, Peru). Earth Surf. Proc. Land. 43, 1373-1389.
\\[-5mm]
\bibitem{} Pudasaini, S.P. (2025): A comprehensive, unified mechanical erosion model for multi-phase mass flows.
Int. J. Multiphase Flow 191, 105328. https://doi.org/10.1016/j.ijmultiphaseflow.2025.105328.
\\[-5mm]
\bibitem{} Pudasaini, S.P., Krautblatter, M. (2021): The mechanics of landslide mobility with erosion. Nat Commun 12, 6793. https://doi.org/10.1038/s41467-021-26959-5.
\\[-5mm]
\bibitem{} Pudasaini, S.P., Fischer, J.-T. (2020): A mechanical model for phase separation in debris flow.
Int. J. Multiphase Flow 129, 103292, https://doi.org/10.1016/j.ijmultiphaseflow.2020.103292.
\\[-5mm]
\bibitem{} Pudasaini, S.P., Mergili, M. (2019): A multi-phase mass flow model. Journal of Geophysical Research:
Earth Surface, 124, 2920-2942.
\\[-5mm]
 \bibitem{} Qiao, C., Ou, G., Pan, H. (2019): Numerical modelling of the long runout character of 2015 Shenzhen landslide with a general two-phase mass flow model. Bull.Eng. Geol. Environ. 78, 3281-3294.
\\[-5mm]
\bibitem{} Santi, P.M., de Wolfe, V.G., Higgins, J.D., Cannon, S.H., Gartner, J.E. (2008): Sources of debris flow material in burned areas. Geomorphology 96, 310-321.
\\[-5mm]
\bibitem{} Somos-Valenzuela, M.A., Chisolm, R.E., Rivas, D.S., Portocarrero, C., McKinney, D.C. (2016): Modeling a glacial lake outburst flood process chain: the case of Lake Palcacocha and Huaraz, Peru. Hydrol. Earth Syst. Sci. 20, 2519-2543.
\\[-5mm]
 \bibitem{} Shen, W., Li, T., Li, P., Berti, M., Shen, Y., Guo, J. (2019): two-layer numerical model for
simulating the frontal plowing phenomenon of flow-like landslides. Engineering Geology 259, 105168.
\\[-5mm]
 \bibitem{}  Theule, J.I., Liebault, F., Laigle, D., Loye, A., Jaboyedoff, M. (2015): Channel scour and fill by debris flows and bedload transport. Geomorphology 243, 92-105.
\\[-5mm]
 \bibitem{} Zhu, K., Gao, L., Zhang, L., Chen, S. (2026): A unified multi-phase debris flow erosion model featuring phase-specific shear stress for internal mechanics and quantifying fine-solid effects. Engineering Geology 372,
108976. https://doi.org/10.1016/j.enggeo.2026.108976.
}

\end{thebibliography}
\end{document}